\documentclass[ floatfix, tightenlines, amsmath, amssymb, aps, prl, lengthcheck, sort&compress, showpacs,  nofootinbib,superscriptaddress]{revtex4-1}

\usepackage{lmodern} 
\usepackage{amsmath}
\usepackage{amssymb}
\usepackage{graphicx}
\usepackage{dsfont}
\usepackage{dcolumn}
\usepackage{units}
\usepackage{wasysym}
\usepackage{multirow}
\usepackage{slashed}
\usepackage[usenames]{color}
\usepackage[colorlinks=true,linkcolor=blue,citecolor=blue,urlcolor=blue]{hyperref}
\usepackage{algpseudocode}
\usepackage{algorithm}
\usepackage{mdframed}
\usepackage{color}
\usepackage{slashed}
\usepackage{xcolor}
\usepackage{booktabs}

\definecolor{violet}{RGB}{111,0,255}
\definecolor{dgreen}{rgb}{0.1,0.50,0.1}
\definecolor{webred}{rgb}{0.75,0,0}

\newcommand{\beq}{\begin{equation}}
\newcommand{\eeq}{\end{equation}}

\def\Slash#1{\setbox0=\hbox{$#1$} 
\dimen0=\wd0 
\setbox1=\hbox{/} \dimen1=\wd1 
\ifdim\dimen0>\dimen1 
\rlap{\hbox to \dimen0{\hfil/\hfil}} 
#1 
\else 
\rlap{\hbox to \dimen1{\hfil$#1$\hfil}} 
/ 
\fi}

\def\longlongrightarrow{\relbar\joinrel\relbar\joinrel\relbar\joinrel\relbar\joinrel\rightarrow}

\begin{document}

\title{On the large-$Q^2$ behavior of the pion transition form factor}

\author{Gernot Eichmann}
\email[e-mail: ]{Gernot.Eichmann@tecnico.ulisboa.pt}
\affiliation{Institut f\"ur Theoretische Physik, Justus-Liebig Universit\"at Gie{\ss}en, 35392 Gie{\ss}en, Germany}
\affiliation{CFTP, Instituto Superior T\'ecnico, Universidade de Lisboa, 1049-001 Lisboa, Portugal}

\author{Christian S. Fischer}
\email[e-mail: ]{Christian.Fischer@physik.uni-giessen.de}
\affiliation{Institut f\"ur Theoretische Physik, Justus-Liebig Universit\"at Gie{\ss}en, 35392 Gie{\ss}en, Germany}
\affiliation{HIC for FAIR Gie{\ss}en, 35392 Gie{\ss}en, Germany}

\author{Esther Weil}

\email[e-mail: ]{Esther.D.Weil@physik.uni-giessen.de}
\affiliation{Institut f\"ur Theoretische Physik, Justus-Liebig Universit\"at Gie{\ss}en, 35392 Gie{\ss}en, Germany}

\author{Richard Williams}
\email[e-mail: ]{Richard.Williams@physik.uni-giessen.de}
\affiliation{Institut f\"ur Theoretische Physik, Justus-Liebig Universit\"at Gie{\ss}en, 35392 Gie{\ss}en, Germany}

\begin{abstract}
We study the transition of non-perturbative to perturbative QCD in situations with possible violations of scaling limits.
To this end we consider the singly- and doubly-virtual pion transition form factor $\pi^0\to\gamma\gamma$ at all momentum scales
of symmetric and asymmetric photon momenta within the Dyson-Schwinger/Bethe-Salpeter approach. 
For the doubly virtual form factor we find good agreement with perturbative asymptotic
scaling laws. For the singly-virtual form factor our results agree with the Belle data. At very large off-shell photon momenta
we identify a mechanism that introduces quantitative modifications to Efremov-Radyushkin-Brodsky-Lepage scaling.

\end{abstract}
\maketitle

The $\pi^0\to\gamma^{(\ast)}\gamma^{(\ast)}$ transition is among the most elementary processes that
allow one to study the evolution between the non-perturbative and the perturbative momentum regions of QCD
\cite{Lepage:1980fj,Efremov:1979qk,Brodsky:1989pv}.
        The transition matrix element reads
        \begin{equation}\label{pigg-current}
            \Lambda^{\mu\nu}(Q,Q') = e^2\,\frac{F(Q^2,{Q'}^2)}{4\pi^2 f_\pi}\,\varepsilon^{\mu\nu\alpha\beta}  {Q'}^\alpha Q^\beta \,,
        \end{equation}
        with incoming and outgoing photon momenta $Q'$ and $Q$, the pion's electroweak decay constant $f_\pi\approx 92$~MeV and the
        electromagnetic charge $e$.
        The process is described by a single transition form factor (TFF) $F(Q^2,{Q'}^2)$, with conventions such that
        $F(0,0)=1$ in the chiral limit due to the Abelian anomaly.

It is a long-standing prediction that for large photon momenta factorization into hard scattering
processes is at work and the TFF reaches the Efremov-Radyush\-kin-Brodsky-Lepage (ERBL) scaling limit~\cite{Lepage:1980fj,Efremov:1979qk}
        \begin{equation}\label{largeQ2-limit}
          \widetilde{F}(Q^2,{Q'}^2) =\frac{\eta_+\, F(Q^2,{Q'}^2)}{4\pi^2 f_\pi^2} \, \stackrel{\eta_+\to\infty}{\longlongrightarrow} \, j(\omega) \,,
        \end{equation}
with $\eta_+ = (Q^2+{Q'}^2)/2$, $\omega=(Q^2-{Q'}^2)/2$ and
        \begin{equation}\label{ERBL}
          j(\omega) = \frac{2}{3}\int_0^1 dx\,\frac{\eta_+^2}{\eta_+^2-\omega^2 (2x-1)^2}\,\varphi_\pi(x)\,.
        \end{equation}
The pion distribution amplitude $\varphi_\pi(x)$ asymptotically approaches $\varphi_\pi(x)\to 6x(1-x)$, so that
$j(0) = \tfrac{2}{3}$ in the symmetric and $j(\pm\eta_+) = 1$ in the asymmetric case.

Whereas this prediction seems to stand on firm ground in the symmetric limit where both photon momenta are asymptotically large,
it has been questioned in the asymmetric limit where one of the photons is on-shell and nonperturbatively interacts with the
pion. Current experimental data on the transition form factor~\cite{Behrend:1990sr,Gronberg:1997fj,Aubert:2009mc,Uehara:2012ag}
indicate that the scale for the onset of the asymptotic behavior could be as large as $10-100 \, \mbox{GeV}^2$, whereas from a
generic factorization picture one would rather expect a scale of order $1\,\mbox{GeV}^2$. Indeed, the situation is not very clear.
Whereas the data from the BaBar collaboration \cite{Aubert:2009mc} seem to indicate that QCD scaling is violated at least for momenta
up to $Q^2 \approx 35 \,\mbox{GeV}^2$, the Belle results \cite{Uehara:2012ag} agree with scaling above $10-15 \, \mbox{GeV}^2$.
The situation may be clarified by upcoming data from BelleII \cite{Abe:2010gxa}.

The potential scaling violations connected to the BaBar data have stirred considerable theoretical
interest in the TFF, see e.g.
\cite{Khodjamirian:1997tk,Anikin:1999cx,Melic:2002ij,Radyushkin:2009zg,Polyakov:2009je,Agaev:2010aq,Agaev:2012tm,Arriola:2010aq,Kroll:2010bf,Gorchtein:2011vf,
Brodsky:2011yv,Brodsky:2011xx,Noguera:2012aw,
ElBennich:2012ij,Dorokhov:2013xpa,Dorokhov:2010zzb,Maris:2002mz,Holl:2005vu,Raya:2015gva,Raya:2016yuj,Mikhailov:2016klg} and references therein.
The results from various theoretical approaches can be classified into three groups \cite{Bakulev:2012nh}:
(i) those that agree with ERBL scaling;
(ii) those that predict a violation of scaling in agreement with the BaBar data;
and (iii) those in between that agree with scaling in principle but maintain a discrepancy to the purely perturbative scaling limit.
Theoretical models that predict scaling deviations typically maintain factorization but employ a
pion distribution amplitude that (strongly) deviates from its expected asymptotic behavior \cite{Agaev:2010aq}.
A different perspective has been advocated in Ref.~\cite{Gorchtein:2011vf}, where resummed gluon exchange diagrams cause
violations of scaling of type (ii).

In this work we identify another mechanism which leads to a deviation of type (iii) at large momenta.
While we exemplify this mechanism at the elementary pion-photon transition process, it is general to all multi-photon processes
with at least one soft photon. Although we use a specific truncation of Dyson-Schwinger equations (DSEs) and Bethe-Salpeter equations (BSEs)
in our calculations, we will also argue that the effect is generic in the sense that it does not depend on the details of the
truncation scheme.

Within the DSE/BSE framework, the large momentum behavior of the TFF has been studied previously in
Refs.~\cite{Raya:2015gva,Raya:2016yuj}. Here we introduce a new technique to compute the TFF on the entire domain of
spacelike momenta. The key novel element, however, that leads to the main result of the present work, is the complete numerical
treatment of the quark-photon vertex including its dynamically generated non-analytic structure associated with vector meson poles.
As it turns out, this structure has a material impact on the large-$Q^2$ behavior of the singly-virtual form factor, leading
to quantitative modifications of ERBL scaling for large momenta.

\begin{figure}[t]
                    \begin{center}
                    \includegraphics[width=0.7\columnwidth]{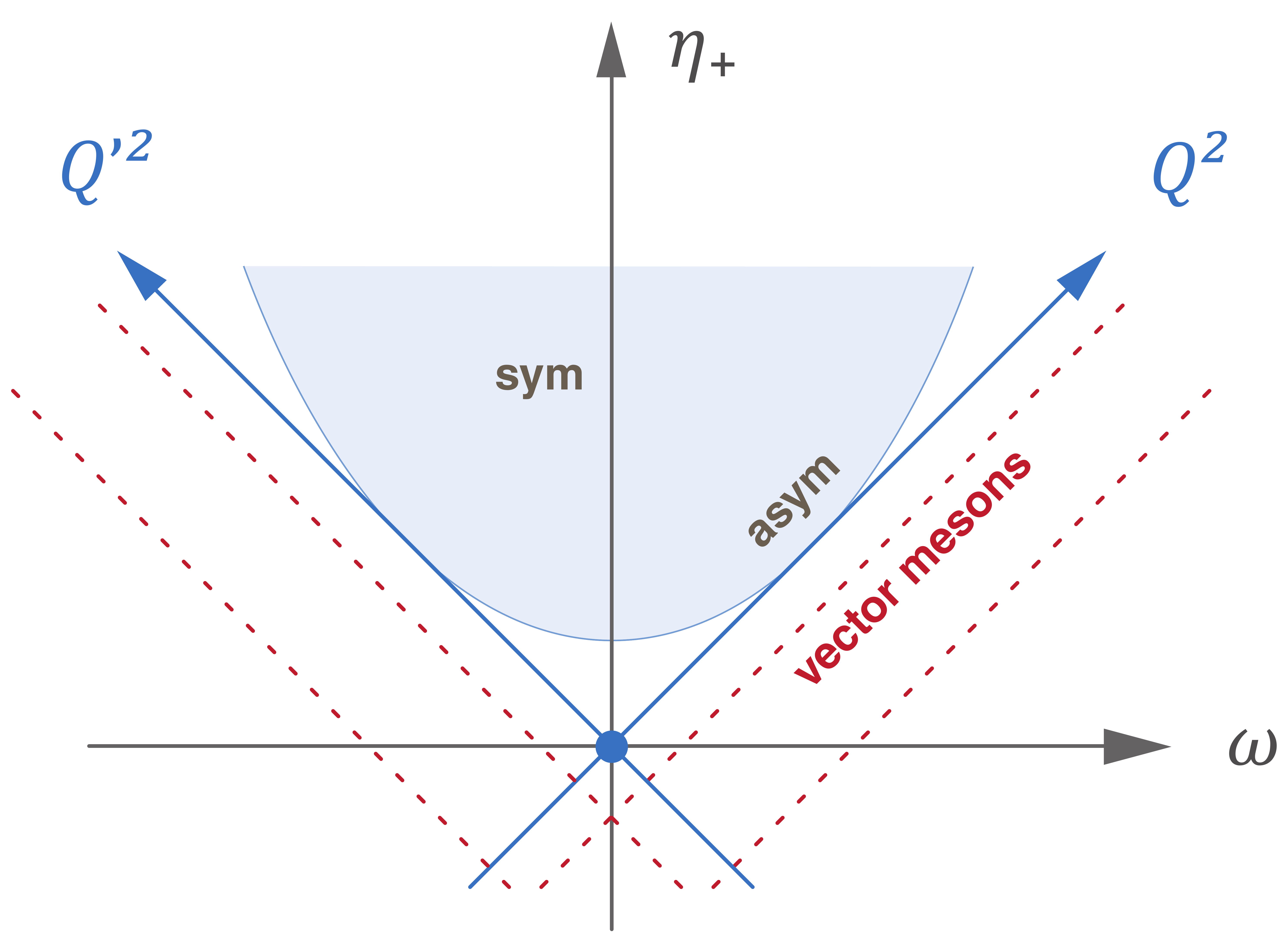}
                    \caption{Kinematic domains in $Q^2$ and ${Q'}^2$  including the symmetric and asymmetric limits.
                            The dotted lines indicate the vector-meson pole locations.
                             The parabola is the spacelike region in the case of constant $t>0$. 
                             }\label{fig:phasespace-1}
                    \end{center}
            \end{figure}

        Our domain of interest is the spacelike region where both $Q^2>0$ and ${Q'}^2>0$, as shown in Fig.~\ref{fig:phasespace-1}.
        It contains the doubly-virtual or symmetric limit $Q^2={Q'}^2$,
        whereas in the singly-virtual or asymmetric case either $Q^2$ or ${Q'}^2$ vanishes.
       The timelike region contains the physical singularities:
       vector-meson poles in the complex plane of $Q^2$ and ${Q'}^2$
            and the corresponding branch cuts from the $\pi\pi$, $K\bar{K}\dots$  continua.

             \begin{figure}[b]
                    \begin{center}
                    \includegraphics[width=1.0\columnwidth]{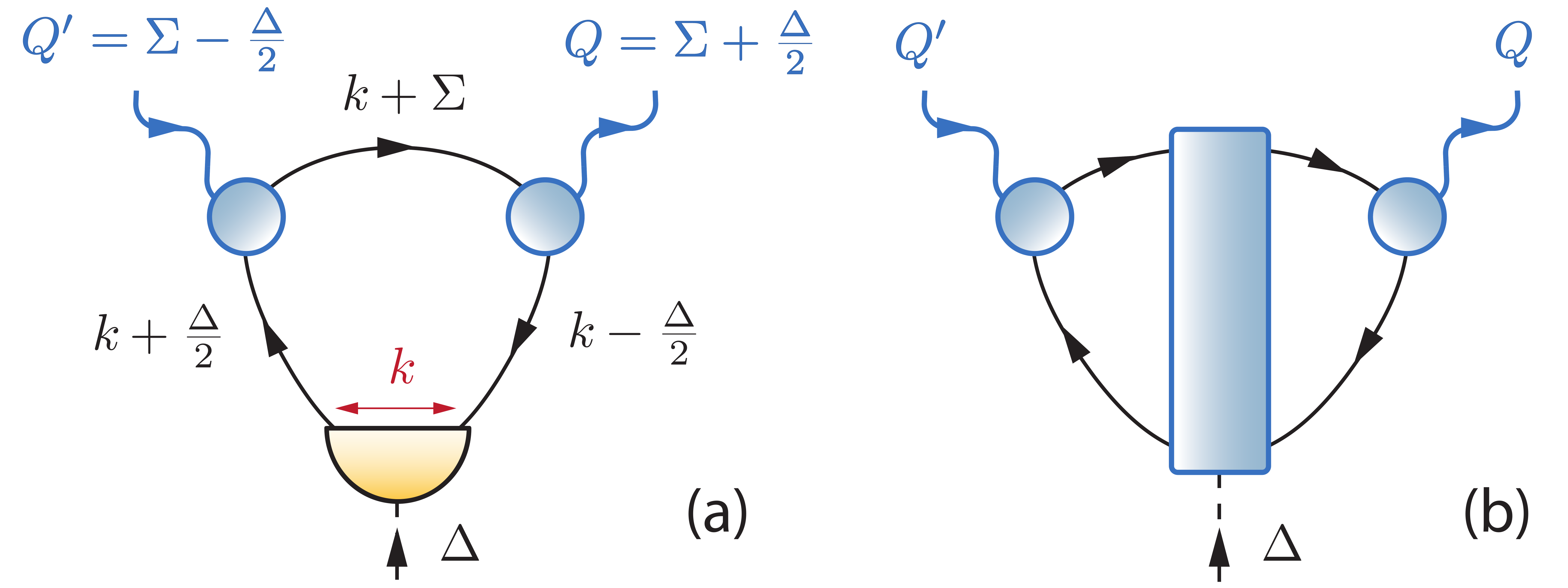}
                    \caption{$\pi^0\to\gamma\gamma$ transition matrix element.
                             }\label{fig:pigg-1}
                    \end{center}
            \end{figure}

        In the following it is useful to work with the average photon momentum $\Sigma^\mu = (Q^\mu + {Q'}^\mu)/2$ and the pion momentum $\Delta^\mu = Q^\mu -{Q'}^\mu$,
        with $\Delta^2  = -m_\pi^2$ for an on-shell pion.
       The two nonperturbative diagrams that constitute the transition matrix element are derived along the lines of Refs.~\cite{Eichmann:2011ec,Eichmann:2012mp} and displayed in Fig.~\ref{fig:pigg-1}.
For the explicit calculations we employ a rainbow-ladder truncation (`Maris-Tandy-model') whose details can be found in Ref.~\cite{Maris:1999bh}.
In that case diagram (b), which contains the pseudoscalar coupling to the $q\bar{q}$ Bethe-Salpeter kernel and thus to the underlying quark-gluon vertex,  does not contribute
and only the triangle diagram (a) survives:
	\begin{equation} \label{eqn:PseudoScalarFormFactor}
    \begin{split}
            \Lambda^{\mu\nu} &= 2e^2\, \text{Tr} \int \!\! \frac{d^4k}{(2\pi)^4} \,  S(k_+)\,\Gamma_\pi(k,\Delta)\,S(k_-)  \\
            &\times \Gamma^\mu(k_-,k+\Sigma)\,S(k+\Sigma)\,\Gamma^\nu(k+\Sigma,k_+) \,,
    \end{split}
	\end{equation}
    where $k_\pm = k \pm \Delta/2$ and
        the color and flavor traces are already worked out.
    The expression depends on three nonperturbative ingredients, which we determine from numerical solutions of their DSEs and BSEs:
the renormalized dressed quark propagator
$S^{-1}(p) = Z_f^{-1}(p^2)\,(i \slashed{p}  + M(p^2))$,
    the pion's Bethe-Salpeter amplitude
\begin{equation}\label{eqn:pion}
\Gamma_\pi(k,\Delta) =  \left(f_1 + f_2\,i\Slash{\Delta}  + f_3\, k\cdot \Delta \,i\Slash{k} +f_4 \left[\Slash{k},\Slash{\Delta}\right]\right) \gamma_5,
\end{equation}
and the dressed quark-photon vertex $\Gamma^\mu(k',k)$.
The quark propagator involves the wave function $Z_f(p^2)$ and the quark mass function $M(p^2)$, which encodes effects
of dynamical mass generation due to the dynamical breaking of chiral symmetry.
The pion amplitude has four components $f_i(k^2,k\cdot \Delta)$,  with $\Delta^2=-m_\pi^2$ fixed. 
The quark-photon vertex can be decomposed into twelve tensors; see e.g. App.~B of Ref.~\cite{Eichmann:2016yit} for details.
Our numerical solution for the vertex
from the inhomogeneous Bethe-Salpeter equation~\cite{Maris:1999bh,Maris:1999ta,Maris:2002mz,Bhagwat:2006pu,Goecke:2010if}
dynamically generates timelike vector-meson poles in its transverse part, so
the underlying physics of vector-meson dominance is
already contained in the form factor without the need for further adjustments.

            \begin{figure}[t]
                    \begin{center}
                    \includegraphics[width=0.6\columnwidth]{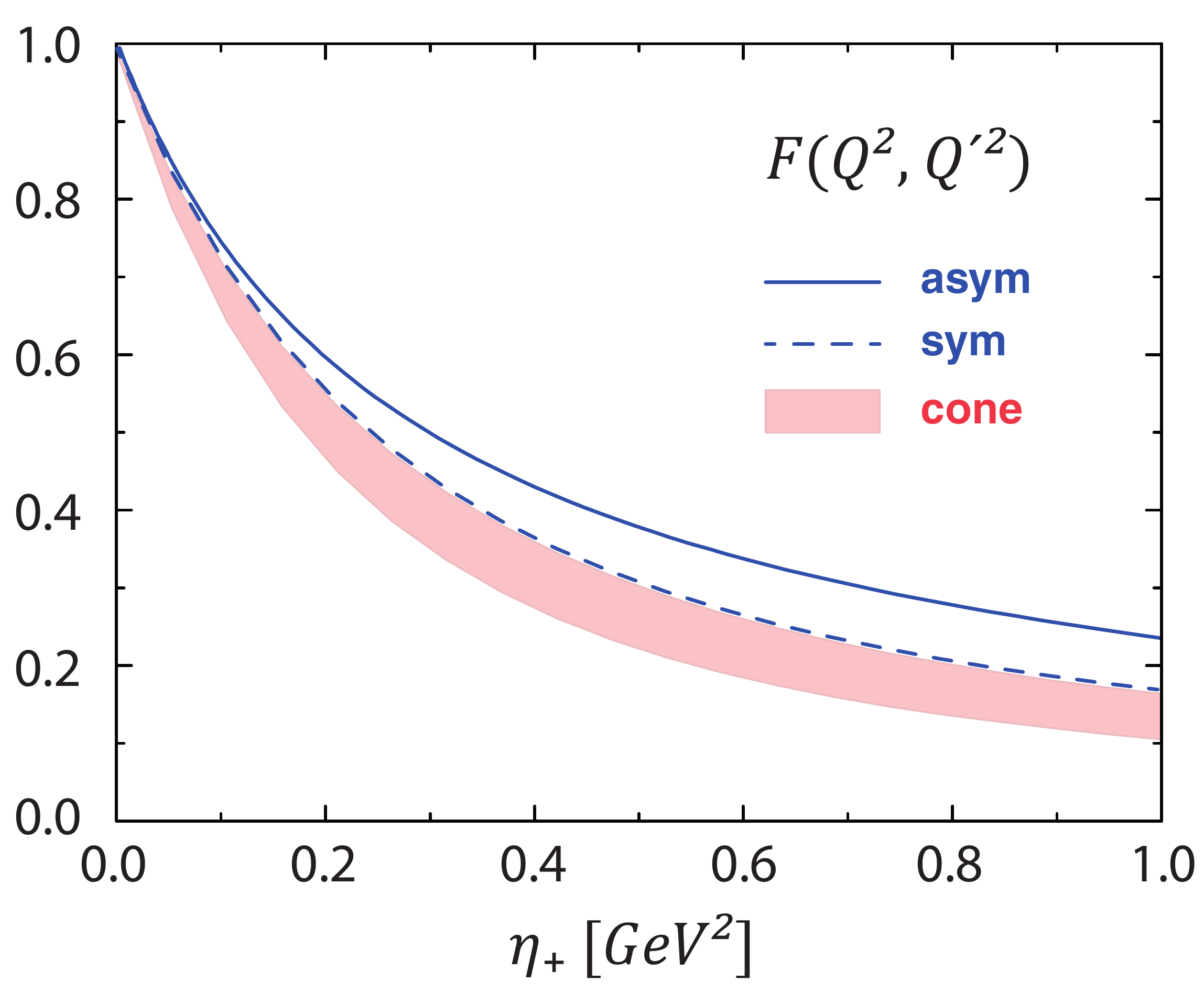}
                    \caption{Onshell transition form factor in the asymmetric (solid) and symmetric limit (dashed)
                             for spacelike momenta $Q^2>0$ and ${Q'}^2>0$. 
                             The band depicts the form factor for all momenta inside the spacelike cone shown in Fig.~\ref{fig:phasespace-2}.
                             }\label{fig:ff-lowQ2}
                    \end{center}
            \end{figure}

In Fig.~\ref{fig:ff-lowQ2} we show the result for the on-shell TFF $F(Q^2,{Q'}^2)$ as a function of the variable $\eta_+$.
The curves for the symmetric and asymmetric limits constitute lower and upper bounds for the TFF in the region $Q^2>0$ and ${Q'}^2>0$. Thus,
for moderate spacelike momenta $\eta_+ \lesssim 1$~GeV$^2$ the TFF mainly scales with $\eta_+$.
The asymptotic result $j(0)=\tfrac{2}{3}$ in the symmetric limit can be recovered from Eq.~\eqref{eqn:PseudoScalarFormFactor},
as explained in~\cite{Maris:2002mz}, and we reproduce it here as well.
Moreover, in the chiral limit the Abelian anomaly entails $F(0,0)=1$ and
our numerical result at the physical pion mass is $F(0,0) = 0.996$.
This actually provides an important consistency check:
replacing both dressed vertices by bare ones would only give $F(0,0) \approx 0.29$;
and even a Ball-Chiu vertex~\cite{Ball:1980ay,Ball:1980ax}, which guarantees charge conservation in the pion's electromagnetic form factor,
produces $F(0,0) \approx 0.86$ only. The transverse structure of the vertex is therefore crucial for a quantitative description of the $\pi^0\to\gamma\gamma$ transition.
An analytical fit function that reproduces our numerical result for the form factor is given elsewhere~\cite{Weil:2017knt}.

Theoretically challenging is the singly-virtual form factor $F(Q^2,0)$;
here an evaluation of Eq.~\eqref{eqn:PseudoScalarFormFactor} is technically difficult because one encounters quark singularities in
the integrand for $Q^2 \gtrsim 4$ GeV$^2$.
             \begin{figure}[t]
                    \begin{center}
                    \includegraphics[width=0.8\columnwidth]{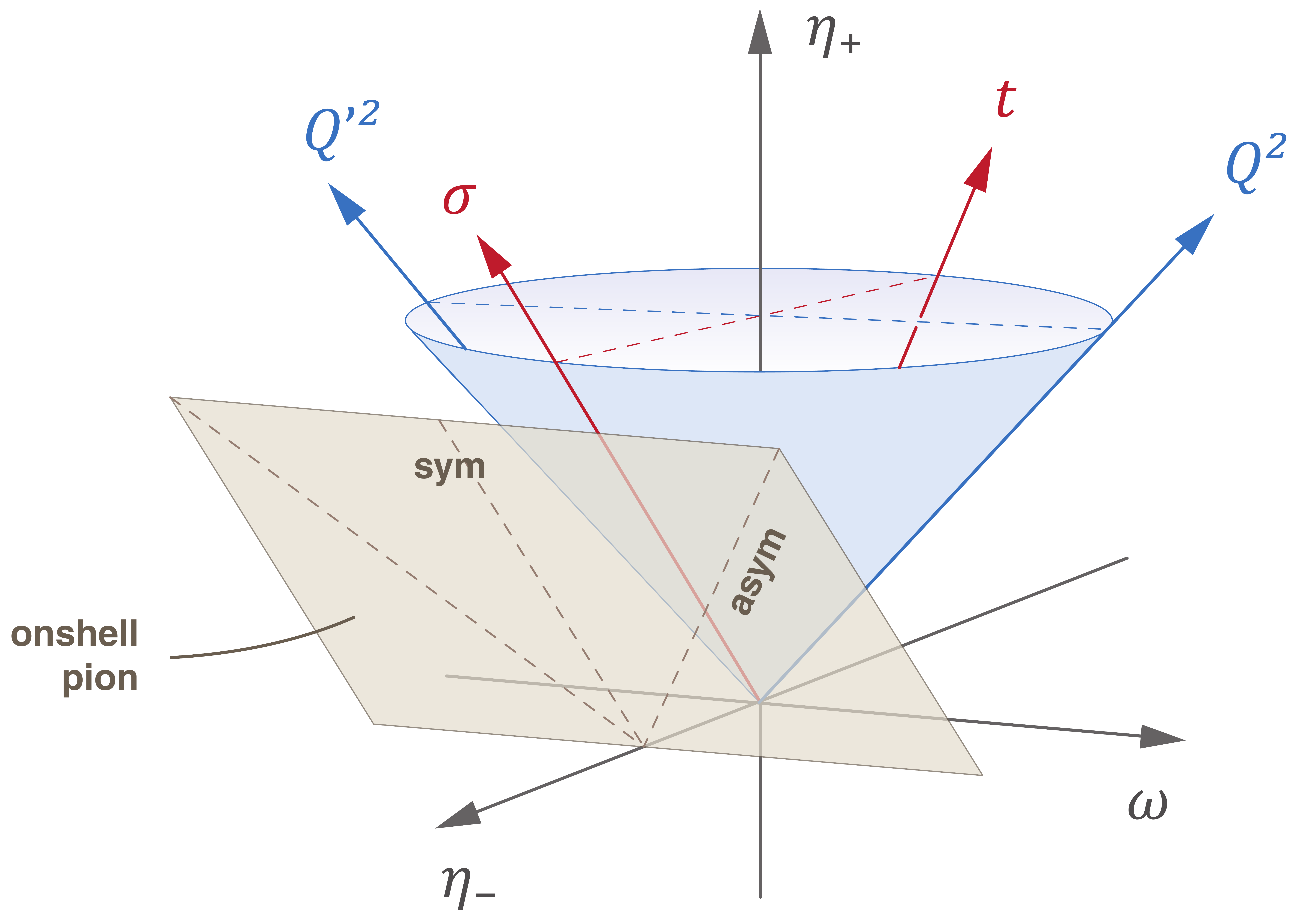}
                    \caption{Three-dimensional view of the $\pi\to\gamma\gamma$ phase space in the variables $\eta_+$, $\eta_-$ and $\omega$.
                             The interior of the cone is the spacelike region for $t>0$, and the plane in front is where the on-shell form factor is defined ($t=-m_\pi^2/4$) as in Fig.~\ref{fig:phasespace-1}.
                             }\label{fig:phasespace-2}
                    \end{center}
            \end{figure}
To overcome the problem, we extract the on-shell form factor from off-shell kinematics based on physical constraints.
These kinematics are important for nucleon Compton scattering or hadronic light-by-light contributions to the anomalous magnetic moment of the muon.
          For the off-shell pion we keep the on-shell dressing functions $f_i$, so that the $\Delta^2$ dependence is carried by the tensor structures alone.
            Denoting $\Sigma^2 = \sigma$ and $\Delta^2=4t$, the transition form factor is then a function of any three of the Lorentz invariants $\{Q^2,\,{Q'}^2,\,Q\cdot Q'\}$,
            $\{\eta_+,\,\eta_-,\,\omega\}$ or $\{\sigma,\,t,\,Z\}$:
            \begin{equation}\label{li-2}   \renewcommand{\arraystretch}{1.2}
            \begin{split}
               \eta_+ &=  \displaystyle \frac{Q^2+{Q'}^2}{2} = \Sigma^2+\frac{\Delta^2}{4} = \sigma +t\,, \\
               \eta_- &= \displaystyle Q\cdot Q' = \Sigma^2-\frac{\Delta^2}{4}  = \sigma - t\,, \\
               \omega &= \displaystyle \frac{Q^2-{Q'}^2}{2} = \Sigma\cdot \Delta = 2\sqrt{\sigma t}\,Z,
            \end{split}
            \end{equation}
            and vice versa: $\{Q^2, {Q'}^2 \} = \eta_+ \pm \omega = \sigma  + t \pm 2\sqrt{\sigma t}\,Z$.

             \begin{figure}[t]
                    \begin{center}
                    \includegraphics[width=0.65\columnwidth]{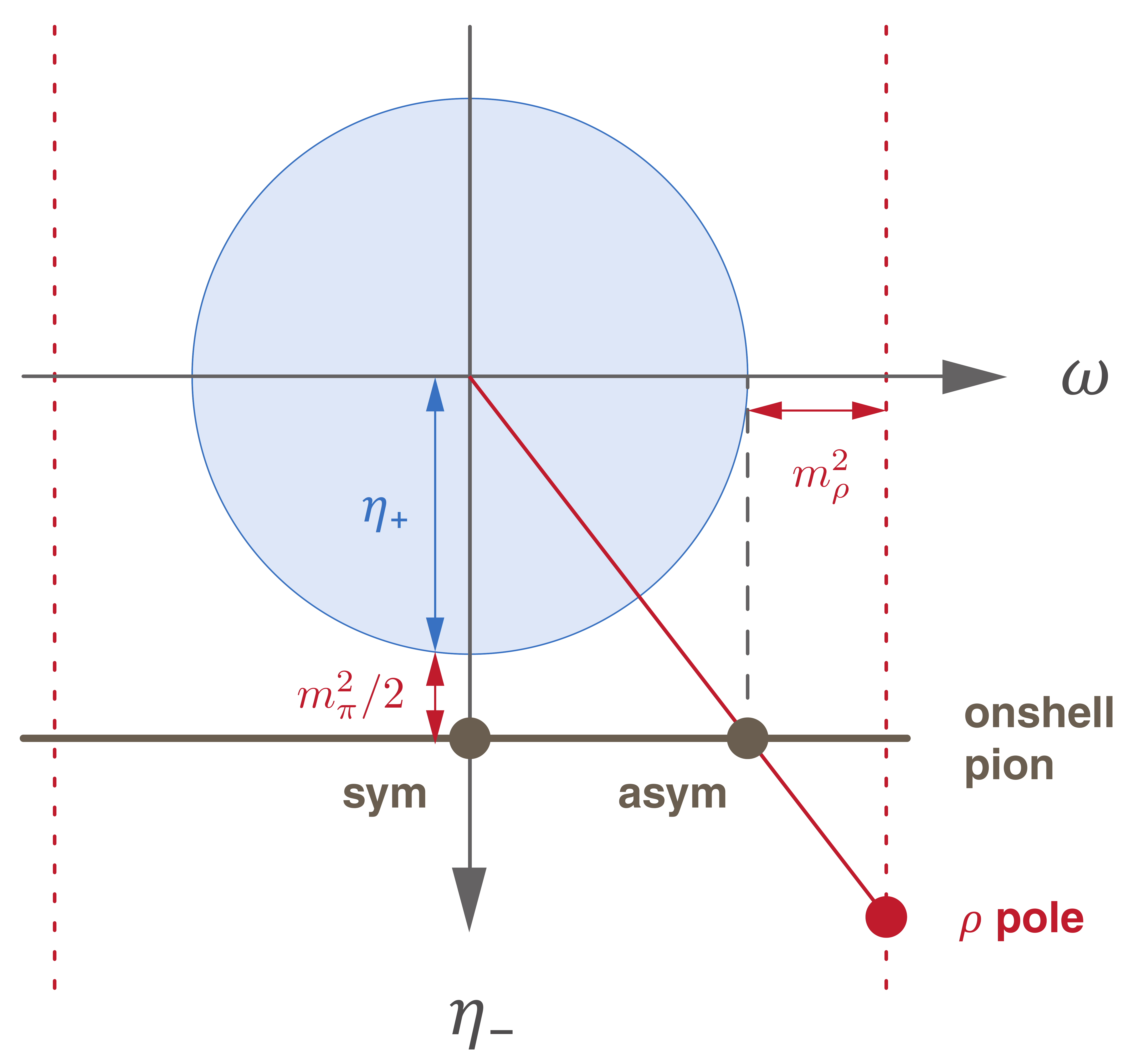}
                    \caption{Same as in Fig.~\ref{fig:phasespace-2} but for fixed $\eta_+$.
                             The arc connecting the origin with the $\rho$ pole
                             determines the singly-virtual form factor.
                             }\label{fig:phasespace-5}
                    \end{center}
                    \begin{center}
                    \includegraphics[width=0.7\columnwidth]{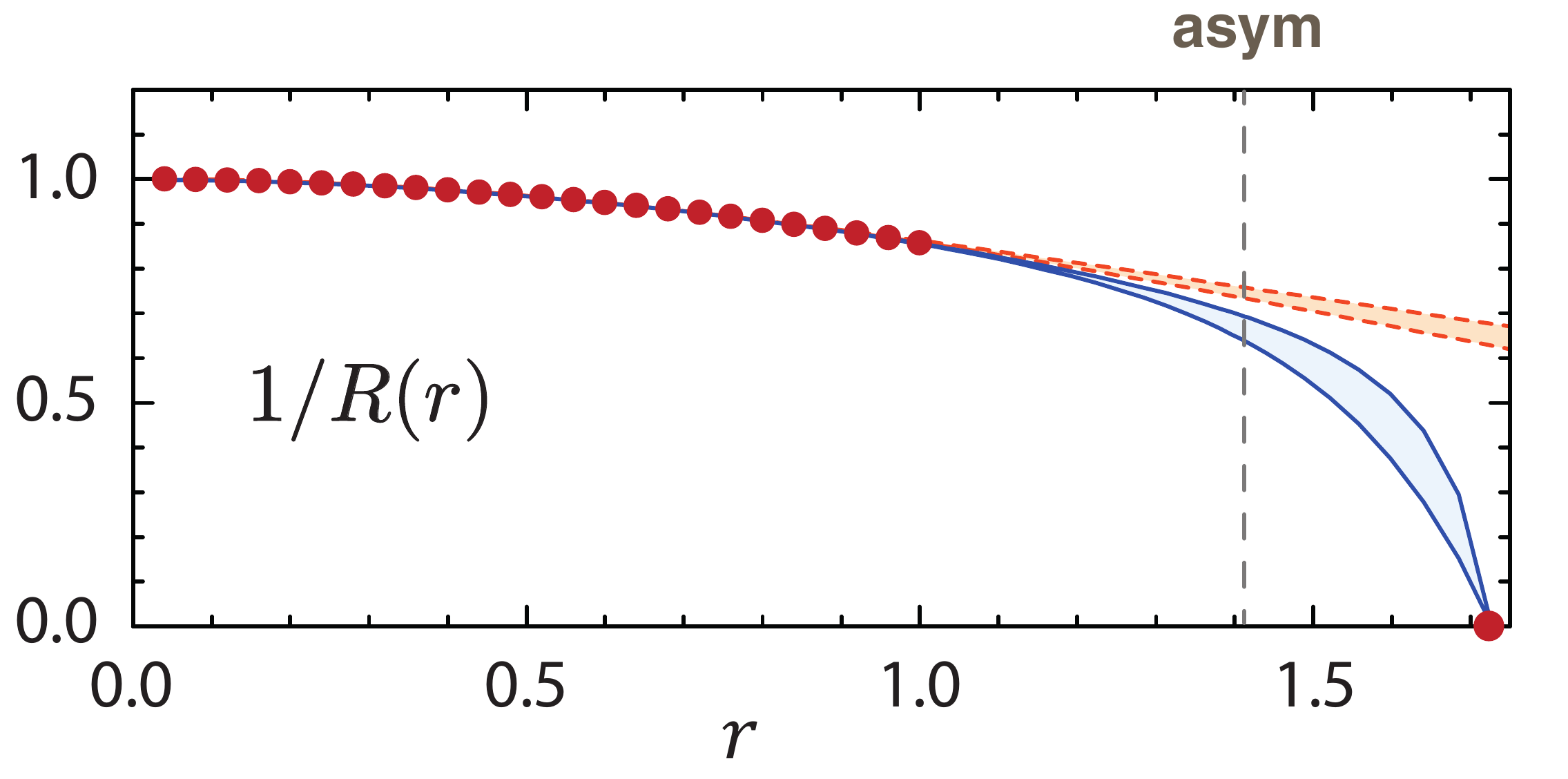}
                    \caption{$1/R(r)$ at  $\eta_+ = 2.5$ GeV$^2$. The points at $r<1$ are calculated
                             and the bands are the fits, once with (solid, blue) and once without (dotted, orange) the constraint at the vector-meson pole.
                             The intercept at $r \approx \sqrt{2}$ determines the singly-virtual on-shell transition form factor.
                             }\label{fig:ff-fit}
                    \end{center}
            \end{figure}

            \begin{figure*}[t]
                    \begin{center}
                    \includegraphics[width=0.78\textwidth]{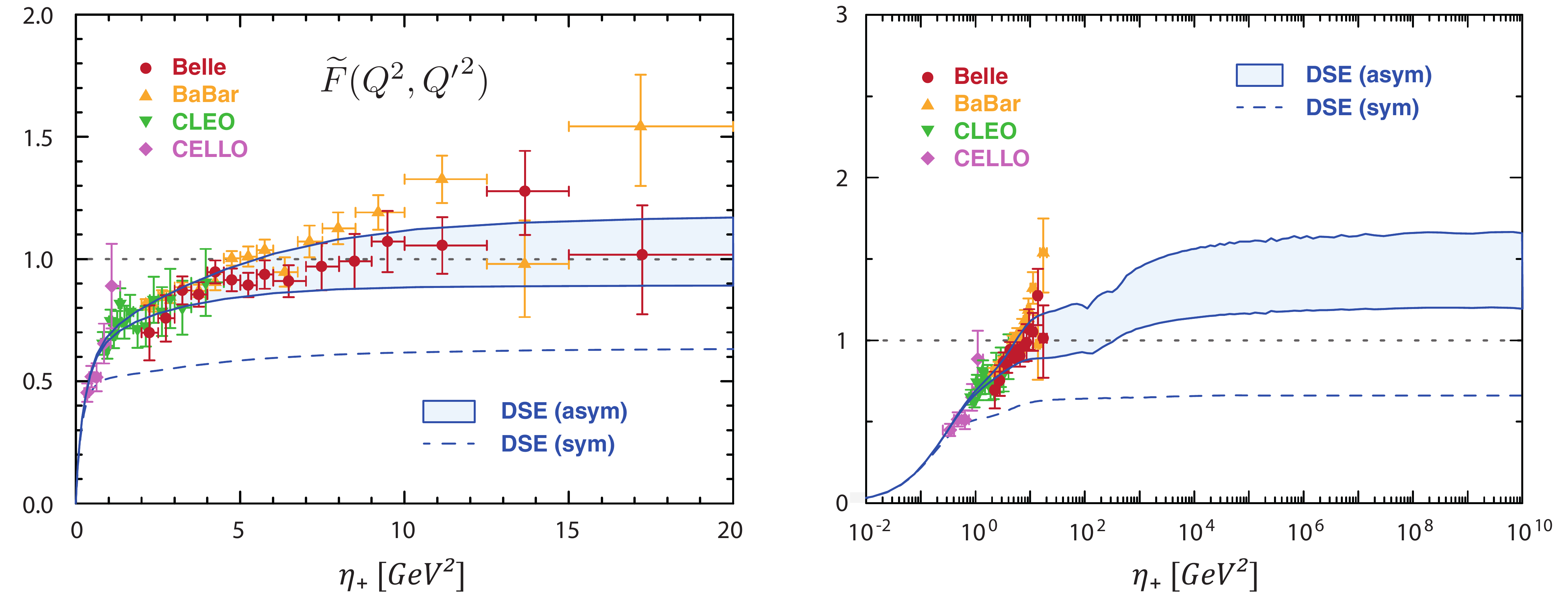}
                    \caption{Weighted on-shell transition form factor defined in Eq.~\eqref{largeQ2-limit} at large $\eta_+$.
                             The band shows the result in the asymmetric limit which is compared to experimental data~\cite{Behrend:1990sr,Gronberg:1997fj,Aubert:2009mc,Uehara:2012ag}, and
                             the dashed line is the form factor in the symmetric limit.
                             The former should asymptotically approach the value 1 (dotted line) whereas the latter converges towards $2/3$.
                             Note that $Q^2=\eta_+$ and $Q^2=2\eta_+$ in the symmetric and asymmetric case, respectively.
                             }\label{fig:ff-largeQ2}
                    \end{center}
            \end{figure*}

               In processes where the photons are also integrated over, $\Sigma^\mu$ becomes the loop momentum and the spacelike region at fixed $t$ is defined by $\sigma > 0$ and $Z \in [-1,1]$,
               which describes the parabola shown in Fig.~\ref{fig:phasespace-1}.
               The conjunction of these parabolas for all possible values of $t>0$ generates a cone
               around the $\eta_+$ axis, which is illustrated in~Fig.~\ref{fig:phasespace-2}.
               The on-shell pion transition current defines a plane  at fixed $t=-m_\pi^2/4$ which,
               for asymptotically large $\eta_+ \gg m_\pi^2$, coincides with the cone in the forward limit $t=0$.

               The interior of the cone is calculable up to arbitrary values of $\eta_+$
               without crossing any singularities in the integrand. The same is also true in the symmetric limit ($\omega=0$) for general $\eta_-$.
Consider then a horizontal plane at some constant value of $\eta_+$ (Fig.~\ref{fig:phasespace-5}).
The circle with radius $\eta_+$ represents the cone. The horizontal line is the on-shell pion plane containing the symmetric and asymmetric limits.
The dashed vertical lines are the nearest vector-meson pole locations.
Since the interaction generates timelike vector-meson poles in the quark-photon vertex and therefore also in the form factor,
the inverse form factor must vanish along these contours.

At a given $\eta_+$ we now consider the quantity
\begin{equation}
    R(r) = \frac{F(\eta_+,\eta_-,\omega)\,}{F(\eta_+,\eta_-,\omega=0)}
\end{equation}
along the arc passing through the asymmetric point.
We divided by the result in the symmetric limit to minimize off-shell momentum dependencies in $\eta_-$.
The radial variable $r$ is defined by $\eta_-^2 + \omega^2 = \eta_+^2 \,r^2$, which for $\eta_+ \gg m^2_\pi$
reduces to $r = \sqrt{2}\,\omega/\eta_+ = \sqrt{2}\,\eta_-/\eta_+$ along the arc.
We calculate~$R(r)$ inside the circle ($r<1$) and employ a fit in the exterior region, which is constrained by
$R(0)=1$ at the center of the circle and
$R(r_v)^{-1}=0$ at the vector-meson pole,  with $r_v = \sqrt{2}\,(1 + m_\rho^2/\eta_+)$.
The fit is illustrated in Fig.~\ref{fig:ff-fit}.
The intersection with the on-shell pion plane ($r = \sqrt{2}$) then determines the transition form factor in the asymmetric limit.
In practice we employ the sum of a polynomial plus a pole term of the form $R(r) = c_0 + c_1 \,r^2 + c_2 \,r^4 + c_3\,(m_\rho^2/\eta_+)/(r_v^2-r^2)$.
The dotted band shows the fit without the pole term, indicating that without the pole constraint the resulting form factor would be substantially smaller.
For $Q^2 \lesssim 4$~GeV$^2$ we confirmed that the fit result coincides with the direct calculation for the TFF in the asymmetric limit.
The procedure can then be repeated for different arcs and different $\eta_+$, thus giving a result for $F(Q^2,{Q'}^2)$ at arbitrary $Q^2>0$ and ${Q'}^2>0$.
            \begin{figure}[b]
                    \begin{center}
                    \includegraphics[width=0.37\textwidth]{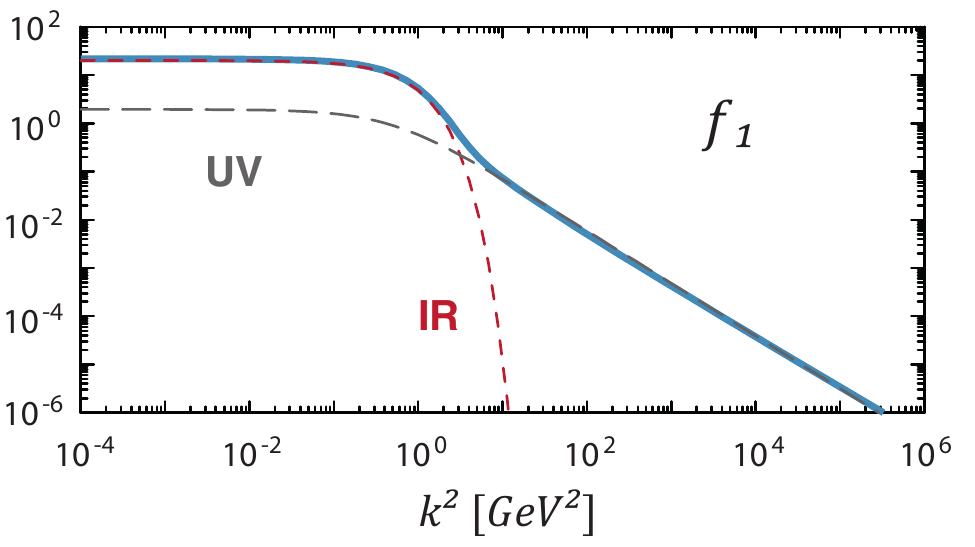}
                    \caption{Leading dressing function $f_1(k^2)$ of the pion Bethe-Salpeter amplitude, Eq.~\eqref{eqn:pion}.
                    The result can be well parametrized by the sum of an exponential in the infrared plus a monopole (with logarithmic corrections) in the UV, as indicated
                    by the dashed lines.
                    }\label{fig:pion}
                    \end{center}
            \end{figure}

The resulting form factor $\widetilde{F}(Q^2,{Q'}^2)$ at large but not asymptotically large momenta  is displayed in the left
panel of Fig.~\ref{fig:ff-largeQ2} and describes the experimental data rather well. It clearly favors the Belle data although
it is also mostly compatible with the BaBar results. We also display the form factor for symmetric photon momenta, which approaches
its asymptotic value~$\tfrac{2}{3}$ already for small values of $\eta_+$. For intermediate values of the photon asymmetry $\omega$,
the TFF is a slowly varying function that monotonically rises from the symmetric to the asymmetric case.

In the right panel of Fig.~\ref{fig:ff-largeQ2} we display the asymmetric form factor at very large momenta. Here we observe a
continued rise beyond the region where experimental data exist. From momenta around $\eta_+ \sim 10^2-10^3$ GeV$^2$ onwards
the form factor settles into its asymptotic behavior, indicating that non-perturbative effects can induce important changes
in observables up to momenta of the order $\sim 30$ GeV. We have explicitly checked that the steep rise in this transition region
is a consequence of the transition from the non-perturbative to the perturbative regime in the leading-order pion amplitude $f_1$
displayed in Fig.~\ref{fig:pion}. It is completely independent of the details of the quark-photon vertex, i.e., we find a qualitatively
similar shape of the TFF if we include $f_1$ only, together with a tree-level propagator and a tree-level quark-photon vertex.

Beyond this transition region the TFF saturates at an asymptotic value, which is quantitatively influenced by the presence of the
$\rho$ pole close to the asymmetric point in Fig.~\ref{fig:phasespace-5} and therefore substantially deviates from the perturbative limit.
Technically, the saturation value is connected to the fit parameter $c_3$, which is almost constant in the momentum range $\eta_+ = 1 \dots 30$ GeV$^2$.
Beyond this point the reach of the vector-meson pole has escaped from the cone and the TFF in its interior is no longer
sensitive to it. Hence we employ that same value as an additional constraint at large $\eta_+$; its variation by a rather conservative $\pm 40\%$ generates the bands in our plots.
Without the physical pole term we recover the ERBL limit~\eqref{ERBL} at asymptotically large $\eta_+$: the fit with a polynomial only yields $\widetilde{F}(\omega = \pm \eta_+) \approx 1$, cf. the dashed orange band in Fig.~\ref{fig:ff-fit}.
Our results therefore imply that in the vicinity of $Q^2=0$ and ${Q'}^2=0$ the TFF is always sensitive to the vector-meson pole, irrespective of
how large $\eta_+$ becomes, and this feature would generate corrections to the ERBL prediction.

We note that the quantitative importance of vector-meson poles up to very large momenta has been realized previously and implemented
in the context of light-cone sum rules, see e.g. \cite{Khodjamirian:1997tk,Agaev:2010aq,Agaev:2012tm,Mikhailov:2016klg}.
There, the fixed-$Q^2$ dispersion relation for the TFF is supplemented with its leading-twist expansion in ${Q'}^2$, based on
quark-hadron duality, which predicts a $1/Q^4$ suppression of the $\rho-$meson pole contribution at large~$Q^2$ and thus ERBL scaling.
The residue of the pole is the $\rho\gamma^\ast\to\pi$ transition matrix element, which in our approach consists of the same diagrams as in Fig.~\ref{fig:pigg-1} 
except that the quark-photon vertex on the left is replaced by the $\rho-$meson Bethe-Salpeter amplitude. For a direct calculation of that matrix element at asymptotic $Q^2$, however,
our present strategy is not applicable since it relies on the lowest-lying pole in the vertex (which is absent from the $\rho-$meson amplitude) 
as a constraint for the interpolation. 
In principle such a suppression could still be enforced in our present context, however only by imposing additional constraints 
on the fit procedure at asymptotically large $Q^2$.  

In any case, the mechanism discussed herein is independent of the approximations used in our calculation, since any realistic truncation of the
underlying quark-gluon interaction has to generate vector-meson poles in the dressed quark-photon vertex. As long as one photon is (close to)
on-shell, the nearest poles are relevant and contribute to the result for the form factor.
The details of the pole locations, i.e. the precise value of its real part or the presence or absence of a width, do not affect the
generic properties at large $\eta_+$. We explicitly checked the quantitative influence of a non-zero, realistic vector-meson width
on the asymptotic value of the form factor and found changes on the level of $1\permil$ as compared to the zero-width result.

In general, we have thus identified a process-independent mechanism that affects all reactions with one external photon
(close to) on-shell and large off-shell momenta from other external legs. 
Whether it truly modifies the ERBL scaling limit or not cannot be reliably determined with the numerical methods at hand;
to further investigate the problem one would need contour-deformation methods (see e.g.~\cite{Weil:2017knt}) 
to directly calculate the relevant integrals at asymptotically large $Q^2$.
This is, however, beyond the scope of this article and relegated to future work.

\paragraph{Acknowledgements}
We thank the referee for very constructive remarks and criticism that helped improve the manuscript.
We are grateful to R.~Alkofer, V.~M.~Braun, S.~Brodsky, P.~Kroll, S.~Leu\-pold and A.~Szczepaniak for enlightening discussions.
This work was supported by the DFG collaborative research centre TR 16, the BMBF grant 05H15RGKBA,
the DFG Project No. FI 970/11-1, the FCT Investigator Grant IF/00898/2015,
the GSI Helmholtzzentrum fuer Schwerionenforschung, and by the Helmholtz International Center for FAIR.

\section*{References}
  \bibliography{baryonspionff} 

\begin{thebibliography}{40}%
\makeatletter
\providecommand \@ifxundefined [1]{%
 \@ifx{#1\undefined}
}%
\providecommand \@ifnum [1]{%
 \ifnum #1\expandafter \@firstoftwo
 \else \expandafter \@secondoftwo
 \fi
}%
\providecommand \@ifx [1]{%
 \ifx #1\expandafter \@firstoftwo
 \else \expandafter \@secondoftwo
 \fi
}%
\providecommand \natexlab [1]{#1}%
\providecommand \enquote  [1]{``#1''}%
\providecommand \bibnamefont  [1]{#1}%
\providecommand \bibfnamefont [1]{#1}%
\providecommand \citenamefont [1]{#1}%
\providecommand \href@noop [0]{\@secondoftwo}%
\providecommand \href [0]{\begingroup \@sanitize@url \@href}%
\providecommand \@href[1]{\@@startlink{#1}\@@href}%
\providecommand \@@href[1]{\endgroup#1\@@endlink}%
\providecommand \@sanitize@url [0]{\catcode `\\12\catcode `\$12\catcode
  `\&12\catcode `\#12\catcode `\^12\catcode `\_12\catcode `\%12\relax}%
\providecommand \@@startlink[1]{}%
\providecommand \@@endlink[0]{}%
\providecommand \url  [0]{\begingroup\@sanitize@url \@url }%
\providecommand \@url [1]{\endgroup\@href {#1}{\urlprefix }}%
\providecommand \urlprefix  [0]{URL }%
\providecommand \Eprint [0]{\href }%
\providecommand \doibase [0]{http://dx.doi.org/}%
\providecommand \selectlanguage [0]{\@gobble}%
\providecommand \bibinfo  [0]{\@secondoftwo}%
\providecommand \bibfield  [0]{\@secondoftwo}%
\providecommand \translation [1]{[#1]}%
\providecommand \BibitemOpen [0]{}%
\providecommand \bibitemStop [0]{}%
\providecommand \bibitemNoStop [0]{.\EOS\space}%
\providecommand \EOS [0]{\spacefactor3000\relax}%
\providecommand \BibitemShut  [1]{\csname bibitem#1\endcsname}%
\let\auto@bib@innerbib\@empty
\bibitem [{\citenamefont {Lepage}\ and\ \citenamefont
  {Brodsky}(1980)}]{Lepage:1980fj}%
  \BibitemOpen
  \bibfield  {author} {\bibinfo {author} {\bibfnamefont {G.~P.}\ \bibnamefont
  {Lepage}}\ and\ \bibinfo {author} {\bibfnamefont {S.~J.}\ \bibnamefont
  {Brodsky}},\ }\href {\doibase 10.1103/PhysRevD.22.2157} {\bibfield  {journal}
  {\bibinfo  {journal} {Phys. Rev.}\ }\textbf {\bibinfo {volume} {D22}},\
  \bibinfo {pages} {2157} (\bibinfo {year} {1980})}\BibitemShut {NoStop}%
\bibitem [{\citenamefont {Efremov}\ and\ \citenamefont
  {Radyushkin}(1980)}]{Efremov:1979qk}%
  \BibitemOpen
  \bibfield  {author} {\bibinfo {author} {\bibfnamefont {A.~V.}\ \bibnamefont
  {Efremov}}\ and\ \bibinfo {author} {\bibfnamefont {A.~V.}\ \bibnamefont
  {Radyushkin}},\ }\href {\doibase 10.1016/0370-2693(80)90869-2} {\bibfield
  {journal} {\bibinfo  {journal} {Phys. Lett.}\ }\textbf {\bibinfo {volume}
  {94B}},\ \bibinfo {pages} {245} (\bibinfo {year} {1980})}\BibitemShut
  {NoStop}%
\bibitem [{\citenamefont {Brodsky}\ and\ \citenamefont
  {Lepage}(1989)}]{Brodsky:1989pv}%
  \BibitemOpen
  \bibfield  {author} {\bibinfo {author} {\bibfnamefont {S.~J.}\ \bibnamefont
  {Brodsky}}\ and\ \bibinfo {author} {\bibfnamefont {G.~P.}\ \bibnamefont
  {Lepage}},\ }\href {\doibase 10.1142/9789814503266_0002} {\bibfield
  {journal} {\bibinfo  {journal} {Adv. Ser. Direct. High Energy Phys.}\
  }\textbf {\bibinfo {volume} {5}},\ \bibinfo {pages} {93} (\bibinfo {year}
  {1989})}\BibitemShut {NoStop}%
\bibitem [{\citenamefont {Behrend}\ \emph {et~al.}(1991)\citenamefont {Behrend}
  \emph {et~al.}}]{Behrend:1990sr}%
  \BibitemOpen
  \bibfield  {author} {\bibinfo {author} {\bibfnamefont {H.~J.}\ \bibnamefont
  {Behrend}} \emph {et~al.} (\bibinfo {collaboration} {CELLO}),\ }\href
  {\doibase 10.1007/BF01549692} {\bibfield  {journal} {\bibinfo  {journal} {Z.
  Phys.}\ }\textbf {\bibinfo {volume} {C49}},\ \bibinfo {pages} {401} (\bibinfo
  {year} {1991})}\BibitemShut {NoStop}%
\bibitem [{\citenamefont {Gronberg}\ \emph {et~al.}(1998)\citenamefont
  {Gronberg} \emph {et~al.}}]{Gronberg:1997fj}%
  \BibitemOpen
  \bibfield  {author} {\bibinfo {author} {\bibfnamefont {J.}~\bibnamefont
  {Gronberg}} \emph {et~al.} (\bibinfo {collaboration} {CLEO}),\ }\href
  {\doibase 10.1103/PhysRevD.57.33} {\bibfield  {journal} {\bibinfo  {journal}
  {Phys. Rev.}\ }\textbf {\bibinfo {volume} {D57}},\ \bibinfo {pages} {33}
  (\bibinfo {year} {1998})},\ \Eprint {http://arxiv.org/abs/hep-ex/9707031}
  {arXiv:hep-ex/9707031 [hep-ex]} \BibitemShut {NoStop}%
\bibitem [{\citenamefont {Aubert}\ \emph {et~al.}(2009)\citenamefont {Aubert}
  \emph {et~al.}}]{Aubert:2009mc}%
  \BibitemOpen
  \bibfield  {author} {\bibinfo {author} {\bibfnamefont {B.}~\bibnamefont
  {Aubert}} \emph {et~al.} (\bibinfo {collaboration} {BaBar}),\ }\href
  {\doibase 10.1103/PhysRevD.80.052002} {\bibfield  {journal} {\bibinfo
  {journal} {Phys. Rev.}\ }\textbf {\bibinfo {volume} {D80}},\ \bibinfo {pages}
  {052002} (\bibinfo {year} {2009})},\ \Eprint {http://arxiv.org/abs/0905.4778}
  {arXiv:0905.4778 [hep-ex]} \BibitemShut {NoStop}%
\bibitem [{\citenamefont {Uehara}\ \emph {et~al.}(2012)\citenamefont {Uehara}
  \emph {et~al.}}]{Uehara:2012ag}%
  \BibitemOpen
  \bibfield  {author} {\bibinfo {author} {\bibfnamefont {S.}~\bibnamefont
  {Uehara}} \emph {et~al.} (\bibinfo {collaboration} {Belle}),\ }\href
  {\doibase 10.1103/PhysRevD.86.092007} {\bibfield  {journal} {\bibinfo
  {journal} {Phys. Rev.}\ }\textbf {\bibinfo {volume} {D86}},\ \bibinfo {pages}
  {092007} (\bibinfo {year} {2012})},\ \Eprint {http://arxiv.org/abs/1205.3249}
  {arXiv:1205.3249 [hep-ex]} \BibitemShut {NoStop}%
\bibitem [{\citenamefont {Abe}\ \emph {et~al.}(2010)\citenamefont {Abe} \emph
  {et~al.}}]{Abe:2010gxa}%
  \BibitemOpen
  \bibfield  {author} {\bibinfo {author} {\bibfnamefont {T.}~\bibnamefont
  {Abe}} \emph {et~al.} (\bibinfo {collaboration} {Belle-II}),\ }\href@noop {}
  {\  (\bibinfo {year} {2010})},\ \Eprint {http://arxiv.org/abs/1011.0352}
  {arXiv:1011.0352 [physics.ins-det]} \BibitemShut {NoStop}%
\bibitem [{\citenamefont {Khodjamirian}(1999)}]{Khodjamirian:1997tk}%
  \BibitemOpen
  \bibfield  {author} {\bibinfo {author} {\bibfnamefont {A.}~\bibnamefont
  {Khodjamirian}},\ }\href {\doibase 10.1007/s100520050357,
  10.1007/s100529800938} {\bibfield  {journal} {\bibinfo  {journal} {Eur. Phys.
  J.}\ }\textbf {\bibinfo {volume} {C6}},\ \bibinfo {pages} {477} (\bibinfo
  {year} {1999})},\ \Eprint {http://arxiv.org/abs/hep-ph/9712451}
  {arXiv:hep-ph/9712451 [hep-ph]} \BibitemShut {NoStop}%
\bibitem [{\citenamefont {Anikin}\ \emph {et~al.}(2000)\citenamefont {Anikin},
  \citenamefont {Dorokhov},\ and\ \citenamefont {Tomio}}]{Anikin:1999cx}%
  \BibitemOpen
  \bibfield  {author} {\bibinfo {author} {\bibfnamefont {I.~V.}\ \bibnamefont
  {Anikin}}, \bibinfo {author} {\bibfnamefont {A.~E.}\ \bibnamefont
  {Dorokhov}}, \ and\ \bibinfo {author} {\bibfnamefont {L.}~\bibnamefont
  {Tomio}},\ }\href {\doibase 10.1016/S0370-2693(00)00097-6} {\bibfield
  {journal} {\bibinfo  {journal} {Phys. Lett.}\ }\textbf {\bibinfo {volume}
  {B475}},\ \bibinfo {pages} {361} (\bibinfo {year} {2000})},\ \Eprint
  {http://arxiv.org/abs/hep-ph/9909368} {arXiv:hep-ph/9909368 [hep-ph]}
  \BibitemShut {NoStop}%
\bibitem [{\citenamefont {Melic}\ \emph {et~al.}(2003)\citenamefont {Melic},
  \citenamefont {Mueller},\ and\ \citenamefont
  {Passek-Kumericki}}]{Melic:2002ij}%
  \BibitemOpen
  \bibfield  {author} {\bibinfo {author} {\bibfnamefont {B.}~\bibnamefont
  {Melic}}, \bibinfo {author} {\bibfnamefont {D.}~\bibnamefont {Mueller}}, \
  and\ \bibinfo {author} {\bibfnamefont {K.}~\bibnamefont {Passek-Kumericki}},\
  }\href {\doibase 10.1103/PhysRevD.68.014013} {\bibfield  {journal} {\bibinfo
  {journal} {Phys. Rev.}\ }\textbf {\bibinfo {volume} {D68}},\ \bibinfo {pages}
  {014013} (\bibinfo {year} {2003})},\ \Eprint
  {http://arxiv.org/abs/hep-ph/0212346} {arXiv:hep-ph/0212346 [hep-ph]}
  \BibitemShut {NoStop}%
\bibitem [{\citenamefont {Radyushkin}(2009)}]{Radyushkin:2009zg}%
  \BibitemOpen
  \bibfield  {author} {\bibinfo {author} {\bibfnamefont {A.~V.}\ \bibnamefont
  {Radyushkin}},\ }\href {\doibase 10.1103/PhysRevD.80.094009} {\bibfield
  {journal} {\bibinfo  {journal} {Phys. Rev.}\ }\textbf {\bibinfo {volume}
  {D80}},\ \bibinfo {pages} {094009} (\bibinfo {year} {2009})},\ \Eprint
  {http://arxiv.org/abs/0906.0323} {arXiv:0906.0323 [hep-ph]} \BibitemShut
  {NoStop}%
\bibitem [{\citenamefont {Polyakov}(2009)}]{Polyakov:2009je}%
  \BibitemOpen
  \bibfield  {author} {\bibinfo {author} {\bibfnamefont {M.~V.}\ \bibnamefont
  {Polyakov}},\ }\href {\doibase 10.1134/S0021364009160024} {\bibfield
  {journal} {\bibinfo  {journal} {JETP Lett.}\ }\textbf {\bibinfo {volume}
  {90}},\ \bibinfo {pages} {228} (\bibinfo {year} {2009})},\ \Eprint
  {http://arxiv.org/abs/0906.0538} {arXiv:0906.0538 [hep-ph]} \BibitemShut
  {NoStop}%
\bibitem [{\citenamefont {Agaev}\ \emph {et~al.}(2011)\citenamefont {Agaev},
  \citenamefont {Braun}, \citenamefont {Offen},\ and\ \citenamefont
  {Porkert}}]{Agaev:2010aq}%
  \BibitemOpen
  \bibfield  {author} {\bibinfo {author} {\bibfnamefont {S.~S.}\ \bibnamefont
  {Agaev}}, \bibinfo {author} {\bibfnamefont {V.~M.}\ \bibnamefont {Braun}},
  \bibinfo {author} {\bibfnamefont {N.}~\bibnamefont {Offen}}, \ and\ \bibinfo
  {author} {\bibfnamefont {F.~A.}\ \bibnamefont {Porkert}},\ }\href {\doibase
  10.1103/PhysRevD.83.054020} {\bibfield  {journal} {\bibinfo  {journal} {Phys.
  Rev.}\ }\textbf {\bibinfo {volume} {D83}},\ \bibinfo {pages} {054020}
  (\bibinfo {year} {2011})},\ \Eprint {http://arxiv.org/abs/1012.4671}
  {arXiv:1012.4671 [hep-ph]} \BibitemShut {NoStop}%
\bibitem [{\citenamefont {Agaev}\ \emph {et~al.}(2012)\citenamefont {Agaev},
  \citenamefont {Braun}, \citenamefont {Offen},\ and\ \citenamefont
  {Porkert}}]{Agaev:2012tm}%
  \BibitemOpen
  \bibfield  {author} {\bibinfo {author} {\bibfnamefont {S.~S.}\ \bibnamefont
  {Agaev}}, \bibinfo {author} {\bibfnamefont {V.~M.}\ \bibnamefont {Braun}},
  \bibinfo {author} {\bibfnamefont {N.}~\bibnamefont {Offen}}, \ and\ \bibinfo
  {author} {\bibfnamefont {F.~A.}\ \bibnamefont {Porkert}},\ }\href {\doibase
  10.1103/PhysRevD.86.077504} {\bibfield  {journal} {\bibinfo  {journal} {Phys.
  Rev.}\ }\textbf {\bibinfo {volume} {D86}},\ \bibinfo {pages} {077504}
  (\bibinfo {year} {2012})},\ \Eprint {http://arxiv.org/abs/1206.3968}
  {arXiv:1206.3968 [hep-ph]} \BibitemShut {NoStop}%
\bibitem [{\citenamefont {Ruiz~Arriola}\ and\ \citenamefont
  {Broniowski}(2010)}]{Arriola:2010aq}%
  \BibitemOpen
  \bibfield  {author} {\bibinfo {author} {\bibfnamefont {E.}~\bibnamefont
  {Ruiz~Arriola}}\ and\ \bibinfo {author} {\bibfnamefont {W.}~\bibnamefont
  {Broniowski}},\ }\href {\doibase 10.1103/PhysRevD.81.094021} {\bibfield
  {journal} {\bibinfo  {journal} {Phys. Rev.}\ }\textbf {\bibinfo {volume}
  {D81}},\ \bibinfo {pages} {094021} (\bibinfo {year} {2010})},\ \Eprint
  {http://arxiv.org/abs/1004.0837} {arXiv:1004.0837 [hep-ph]} \BibitemShut
  {NoStop}%
\bibitem [{\citenamefont {Kroll}(2011)}]{Kroll:2010bf}%
  \BibitemOpen
  \bibfield  {author} {\bibinfo {author} {\bibfnamefont {P.}~\bibnamefont
  {Kroll}},\ }\href {\doibase 10.1140/epjc/s10052-011-1623-4} {\bibfield
  {journal} {\bibinfo  {journal} {Eur. Phys. J.}\ }\textbf {\bibinfo {volume}
  {C71}},\ \bibinfo {pages} {1623} (\bibinfo {year} {2011})},\ \Eprint
  {http://arxiv.org/abs/1012.3542} {arXiv:1012.3542 [hep-ph]} \BibitemShut
  {NoStop}%
\bibitem [{\citenamefont {Gorchtein}\ \emph {et~al.}(2012)\citenamefont
  {Gorchtein}, \citenamefont {Guo},\ and\ \citenamefont
  {Szczepaniak}}]{Gorchtein:2011vf}%
  \BibitemOpen
  \bibfield  {author} {\bibinfo {author} {\bibfnamefont {M.}~\bibnamefont
  {Gorchtein}}, \bibinfo {author} {\bibfnamefont {P.}~\bibnamefont {Guo}}, \
  and\ \bibinfo {author} {\bibfnamefont {A.~P.}\ \bibnamefont {Szczepaniak}},\
  }\href {\doibase 10.1103/PhysRevC.86.015205} {\bibfield  {journal} {\bibinfo
  {journal} {Phys. Rev.}\ }\textbf {\bibinfo {volume} {C86}},\ \bibinfo {pages}
  {015205} (\bibinfo {year} {2012})},\ \Eprint {http://arxiv.org/abs/1102.5558}
  {arXiv:1102.5558 [nucl-th]} \BibitemShut {NoStop}%
\bibitem [{\citenamefont {Brodsky}\ \emph
  {et~al.}(2011{\natexlab{a}})\citenamefont {Brodsky}, \citenamefont {Cao},\
  and\ \citenamefont {de~Teramond}}]{Brodsky:2011yv}%
  \BibitemOpen
  \bibfield  {author} {\bibinfo {author} {\bibfnamefont {S.~J.}\ \bibnamefont
  {Brodsky}}, \bibinfo {author} {\bibfnamefont {F.-G.}\ \bibnamefont {Cao}}, \
  and\ \bibinfo {author} {\bibfnamefont {G.~F.}\ \bibnamefont {de~Teramond}},\
  }\href {\doibase 10.1103/PhysRevD.84.033001} {\bibfield  {journal} {\bibinfo
  {journal} {Phys. Rev.}\ }\textbf {\bibinfo {volume} {D84}},\ \bibinfo {pages}
  {033001} (\bibinfo {year} {2011}{\natexlab{a}})},\ \Eprint
  {http://arxiv.org/abs/1104.3364} {arXiv:1104.3364 [hep-ph]} \BibitemShut
  {NoStop}%
\bibitem [{\citenamefont {Brodsky}\ \emph
  {et~al.}(2011{\natexlab{b}})\citenamefont {Brodsky}, \citenamefont {Cao},\
  and\ \citenamefont {de~Teramond}}]{Brodsky:2011xx}%
  \BibitemOpen
  \bibfield  {author} {\bibinfo {author} {\bibfnamefont {S.~J.}\ \bibnamefont
  {Brodsky}}, \bibinfo {author} {\bibfnamefont {F.-G.}\ \bibnamefont {Cao}}, \
  and\ \bibinfo {author} {\bibfnamefont {G.~F.}\ \bibnamefont {de~Teramond}},\
  }\href {\doibase 10.1103/PhysRevD.84.075012} {\bibfield  {journal} {\bibinfo
  {journal} {Phys. Rev.}\ }\textbf {\bibinfo {volume} {D84}},\ \bibinfo {pages}
  {075012} (\bibinfo {year} {2011}{\natexlab{b}})},\ \Eprint
  {http://arxiv.org/abs/1105.3999} {arXiv:1105.3999 [hep-ph]} \BibitemShut
  {NoStop}%
\bibitem [{\citenamefont {Noguera}\ and\ \citenamefont
  {Vento}(2012)}]{Noguera:2012aw}%
  \BibitemOpen
  \bibfield  {author} {\bibinfo {author} {\bibfnamefont {S.}~\bibnamefont
  {Noguera}}\ and\ \bibinfo {author} {\bibfnamefont {V.}~\bibnamefont
  {Vento}},\ }\href {\doibase 10.1140/epja/i2012-12143-1} {\bibfield  {journal}
  {\bibinfo  {journal} {Eur. Phys. J.}\ }\textbf {\bibinfo {volume} {A48}},\
  \bibinfo {pages} {143} (\bibinfo {year} {2012})},\ \Eprint
  {http://arxiv.org/abs/1205.4598} {arXiv:1205.4598 [hep-ph]} \BibitemShut
  {NoStop}%
\bibitem [{\citenamefont {El-Bennich}\ \emph {et~al.}(2013)\citenamefont
  {El-Bennich}, \citenamefont {de~Melo},\ and\ \citenamefont
  {Frederico}}]{ElBennich:2012ij}%
  \BibitemOpen
  \bibfield  {author} {\bibinfo {author} {\bibfnamefont {B.}~\bibnamefont
  {El-Bennich}}, \bibinfo {author} {\bibfnamefont {J.~P. B.~C.}\ \bibnamefont
  {de~Melo}}, \ and\ \bibinfo {author} {\bibfnamefont {T.}~\bibnamefont
  {Frederico}},\ }\href {\doibase 10.1007/s00601-013-0682-5} {\bibfield
  {journal} {\bibinfo  {journal} {Few Body Syst.}\ }\textbf {\bibinfo {volume}
  {54}},\ \bibinfo {pages} {1851} (\bibinfo {year} {2013})},\ \Eprint
  {http://arxiv.org/abs/1211.2829} {arXiv:1211.2829 [nucl-th]} \BibitemShut
  {NoStop}%
\bibitem [{\citenamefont {Dorokhov}\ and\ \citenamefont
  {Kuraev}(2013)}]{Dorokhov:2013xpa}%
  \BibitemOpen
  \bibfield  {author} {\bibinfo {author} {\bibfnamefont {A.~E.}\ \bibnamefont
  {Dorokhov}}\ and\ \bibinfo {author} {\bibfnamefont {E.~A.}\ \bibnamefont
  {Kuraev}},\ }\href {\doibase 10.1103/PhysRevD.88.014038} {\bibfield
  {journal} {\bibinfo  {journal} {Phys. Rev.}\ }\textbf {\bibinfo {volume}
  {D88}},\ \bibinfo {pages} {014038} (\bibinfo {year} {2013})},\ \Eprint
  {http://arxiv.org/abs/1305.0888} {arXiv:1305.0888 [hep-ph]} \BibitemShut
  {NoStop}%
\bibitem [{\citenamefont {Dorokhov}(2010)}]{Dorokhov:2010zzb}%
  \BibitemOpen
  \bibfield  {author} {\bibinfo {author} {\bibfnamefont {A.~E.}\ \bibnamefont
  {Dorokhov}},\ }\href {\doibase 10.1134/S0021364010220145} {\bibfield
  {journal} {\bibinfo  {journal} {JETP Lett.}\ }\textbf {\bibinfo {volume}
  {92}},\ \bibinfo {pages} {707} (\bibinfo {year} {2010})}\BibitemShut
  {NoStop}%
\bibitem [{\citenamefont {Maris}\ and\ \citenamefont
  {Tandy}(2002)}]{Maris:2002mz}%
  \BibitemOpen
  \bibfield  {author} {\bibinfo {author} {\bibfnamefont {P.}~\bibnamefont
  {Maris}}\ and\ \bibinfo {author} {\bibfnamefont {P.~C.}\ \bibnamefont
  {Tandy}},\ }\href {\doibase 10.1103/PhysRevC.65.045211} {\bibfield  {journal}
  {\bibinfo  {journal} {Phys. Rev.}\ }\textbf {\bibinfo {volume} {C65}},\
  \bibinfo {pages} {045211} (\bibinfo {year} {2002})},\ \Eprint
  {http://arxiv.org/abs/nucl-th/0201017} {arXiv:nucl-th/0201017 [nucl-th]}
  \BibitemShut {NoStop}%
\bibitem [{\citenamefont {Holl}\ \emph {et~al.}(2005)\citenamefont {Holl},
  \citenamefont {Krassnigg}, \citenamefont {Maris}, \citenamefont {Roberts},\
  and\ \citenamefont {Wright}}]{Holl:2005vu}%
  \BibitemOpen
  \bibfield  {author} {\bibinfo {author} {\bibfnamefont {A.}~\bibnamefont
  {Holl}}, \bibinfo {author} {\bibfnamefont {A.}~\bibnamefont {Krassnigg}},
  \bibinfo {author} {\bibfnamefont {P.}~\bibnamefont {Maris}}, \bibinfo
  {author} {\bibfnamefont {C.~D.}\ \bibnamefont {Roberts}}, \ and\ \bibinfo
  {author} {\bibfnamefont {S.~V.}\ \bibnamefont {Wright}},\ }\href {\doibase
  10.1103/PhysRevC.71.065204} {\bibfield  {journal} {\bibinfo  {journal} {Phys.
  Rev.}\ }\textbf {\bibinfo {volume} {C71}},\ \bibinfo {pages} {065204}
  (\bibinfo {year} {2005})},\ \Eprint {http://arxiv.org/abs/nucl-th/0503043}
  {arXiv:nucl-th/0503043 [nucl-th]} \BibitemShut {NoStop}%
\bibitem [{\citenamefont {Raya}\ \emph
  {et~al.}(2016{\natexlab{a}})\citenamefont {Raya}, \citenamefont {Chang},
  \citenamefont {Bashir}, \citenamefont {Cobos-Martinez}, \citenamefont
  {Gutierrez-Guerrero}, \citenamefont {Roberts},\ and\ \citenamefont
  {Tandy}}]{Raya:2015gva}%
  \BibitemOpen
  \bibfield  {author} {\bibinfo {author} {\bibfnamefont {K.}~\bibnamefont
  {Raya}}, \bibinfo {author} {\bibfnamefont {L.}~\bibnamefont {Chang}},
  \bibinfo {author} {\bibfnamefont {A.}~\bibnamefont {Bashir}}, \bibinfo
  {author} {\bibfnamefont {J.~J.}\ \bibnamefont {Cobos-Martinez}}, \bibinfo
  {author} {\bibfnamefont {L.~X.}\ \bibnamefont {Gutierrez-Guerrero}}, \bibinfo
  {author} {\bibfnamefont {C.~D.}\ \bibnamefont {Roberts}}, \ and\ \bibinfo
  {author} {\bibfnamefont {P.~C.}\ \bibnamefont {Tandy}},\ }\href {\doibase
  10.1103/PhysRevD.93.074017} {\bibfield  {journal} {\bibinfo  {journal} {Phys.
  Rev.}\ }\textbf {\bibinfo {volume} {D93}},\ \bibinfo {pages} {074017}
  (\bibinfo {year} {2016}{\natexlab{a}})},\ \Eprint
  {http://arxiv.org/abs/1510.02799} {arXiv:1510.02799 [nucl-th]} \BibitemShut
  {NoStop}%
\bibitem [{\citenamefont {Raya}\ \emph
  {et~al.}(2016{\natexlab{b}})\citenamefont {Raya}, \citenamefont {Ding},
  \citenamefont {Bashir}, \citenamefont {Chang},\ and\ \citenamefont
  {Roberts}}]{Raya:2016yuj}%
  \BibitemOpen
  \bibfield  {author} {\bibinfo {author} {\bibfnamefont {K.}~\bibnamefont
  {Raya}}, \bibinfo {author} {\bibfnamefont {M.}~\bibnamefont {Ding}}, \bibinfo
  {author} {\bibfnamefont {A.}~\bibnamefont {Bashir}}, \bibinfo {author}
  {\bibfnamefont {L.}~\bibnamefont {Chang}}, \ and\ \bibinfo {author}
  {\bibfnamefont {C.~D.}\ \bibnamefont {Roberts}},\ }\href@noop {} {\
  (\bibinfo {year} {2016}{\natexlab{b}})},\ \Eprint
  {http://arxiv.org/abs/1610.06575} {arXiv:1610.06575 [nucl-th]} \BibitemShut
  {NoStop}%
\bibitem [{\citenamefont {Mikhailov}\ \emph {et~al.}(2016)\citenamefont
  {Mikhailov}, \citenamefont {Pimikov},\ and\ \citenamefont
  {Stefanis}}]{Mikhailov:2016klg}%
  \BibitemOpen
  \bibfield  {author} {\bibinfo {author} {\bibfnamefont {S.~V.}\ \bibnamefont
  {Mikhailov}}, \bibinfo {author} {\bibfnamefont {A.~V.}\ \bibnamefont
  {Pimikov}}, \ and\ \bibinfo {author} {\bibfnamefont {N.~G.}\ \bibnamefont
  {Stefanis}},\ }\href {\doibase 10.1103/PhysRevD.93.114018} {\bibfield
  {journal} {\bibinfo  {journal} {Phys. Rev.}\ }\textbf {\bibinfo {volume}
  {D93}},\ \bibinfo {pages} {114018} (\bibinfo {year} {2016})},\ \Eprint
  {http://arxiv.org/abs/1604.06391} {arXiv:1604.06391 [hep-ph]} \BibitemShut
  {NoStop}%
\bibitem [{\citenamefont {Bakulev}\ \emph {et~al.}(2012)\citenamefont
  {Bakulev}, \citenamefont {Mikhailov}, \citenamefont {Pimikov},\ and\
  \citenamefont {Stefanis}}]{Bakulev:2012nh}%
  \BibitemOpen
  \bibfield  {author} {\bibinfo {author} {\bibfnamefont {A.~P.}\ \bibnamefont
  {Bakulev}}, \bibinfo {author} {\bibfnamefont {S.~V.}\ \bibnamefont
  {Mikhailov}}, \bibinfo {author} {\bibfnamefont {A.~V.}\ \bibnamefont
  {Pimikov}}, \ and\ \bibinfo {author} {\bibfnamefont {N.~G.}\ \bibnamefont
  {Stefanis}},\ }\href {\doibase 10.1103/PhysRevD.86.031501} {\bibfield
  {journal} {\bibinfo  {journal} {Phys. Rev.}\ }\textbf {\bibinfo {volume}
  {D86}},\ \bibinfo {pages} {031501} (\bibinfo {year} {2012})},\ \Eprint
  {http://arxiv.org/abs/1205.3770} {arXiv:1205.3770 [hep-ph]} \BibitemShut
  {NoStop}%
\bibitem [{\citenamefont {Eichmann}\ and\ \citenamefont
  {Fischer}(2012)}]{Eichmann:2011ec}%
  \BibitemOpen
  \bibfield  {author} {\bibinfo {author} {\bibfnamefont {G.}~\bibnamefont
  {Eichmann}}\ and\ \bibinfo {author} {\bibfnamefont {C.~S.}\ \bibnamefont
  {Fischer}},\ }\href {\doibase 10.1103/PhysRevD.85.034015} {\bibfield
  {journal} {\bibinfo  {journal} {Phys. Rev.}\ }\textbf {\bibinfo {volume}
  {D85}},\ \bibinfo {pages} {034015} (\bibinfo {year} {2012})},\ \Eprint
  {http://arxiv.org/abs/1111.0197} {arXiv:1111.0197 [hep-ph]} \BibitemShut
  {NoStop}%
\bibitem [{\citenamefont {Eichmann}\ and\ \citenamefont
  {Fischer}(2013)}]{Eichmann:2012mp}%
  \BibitemOpen
  \bibfield  {author} {\bibinfo {author} {\bibfnamefont {G.}~\bibnamefont
  {Eichmann}}\ and\ \bibinfo {author} {\bibfnamefont {C.~S.}\ \bibnamefont
  {Fischer}},\ }\href {\doibase 10.1103/PhysRevD.87.036006} {\bibfield
  {journal} {\bibinfo  {journal} {Phys. Rev.}\ }\textbf {\bibinfo {volume}
  {D87}},\ \bibinfo {pages} {036006} (\bibinfo {year} {2013})},\ \Eprint
  {http://arxiv.org/abs/1212.1761} {arXiv:1212.1761 [hep-ph]} \BibitemShut
  {NoStop}%
\bibitem [{\citenamefont {Maris}\ and\ \citenamefont
  {Tandy}(2000{\natexlab{a}})}]{Maris:1999bh}%
  \BibitemOpen
  \bibfield  {author} {\bibinfo {author} {\bibfnamefont {P.}~\bibnamefont
  {Maris}}\ and\ \bibinfo {author} {\bibfnamefont {P.~C.}\ \bibnamefont
  {Tandy}},\ }\href {\doibase 10.1103/PhysRevC.61.045202} {\bibfield  {journal}
  {\bibinfo  {journal} {Phys. Rev.}\ }\textbf {\bibinfo {volume} {C61}},\
  \bibinfo {pages} {045202} (\bibinfo {year} {2000}{\natexlab{a}})},\ \Eprint
  {http://arxiv.org/abs/nucl-th/9910033} {arXiv:nucl-th/9910033 [nucl-th]}
  \BibitemShut {NoStop}%
\bibitem [{\citenamefont {Eichmann}\ \emph {et~al.}(2016)\citenamefont
  {Eichmann}, \citenamefont {Sanchis-Alepuz}, \citenamefont {Williams},
  \citenamefont {Alkofer},\ and\ \citenamefont {Fischer}}]{Eichmann:2016yit}%
  \BibitemOpen
  \bibfield  {author} {\bibinfo {author} {\bibfnamefont {G.}~\bibnamefont
  {Eichmann}}, \bibinfo {author} {\bibfnamefont {H.}~\bibnamefont
  {Sanchis-Alepuz}}, \bibinfo {author} {\bibfnamefont {R.}~\bibnamefont
  {Williams}}, \bibinfo {author} {\bibfnamefont {R.}~\bibnamefont {Alkofer}}, \
  and\ \bibinfo {author} {\bibfnamefont {C.~S.}\ \bibnamefont {Fischer}},\
  }\href {\doibase 10.1016/j.ppnp.2016.07.001} {\bibfield  {journal} {\bibinfo
  {journal} {Prog. Part. Nucl. Phys.}\ }\textbf {\bibinfo {volume} {91}},\
  \bibinfo {pages} {1} (\bibinfo {year} {2016})},\ \Eprint
  {http://arxiv.org/abs/1606.09602} {arXiv:1606.09602 [hep-ph]} \BibitemShut
  {NoStop}%
\bibitem [{\citenamefont {Maris}\ and\ \citenamefont
  {Tandy}(2000{\natexlab{b}})}]{Maris:1999ta}%
  \BibitemOpen
  \bibfield  {author} {\bibinfo {author} {\bibfnamefont {P.}~\bibnamefont
  {Maris}}\ and\ \bibinfo {author} {\bibfnamefont {P.~C.}\ \bibnamefont
  {Tandy}},\ }\href {\doibase 10.1016/S0375-9474(99)00627-2} {\bibfield
  {journal} {\bibinfo  {journal} {Nucl. Phys.}\ }\textbf {\bibinfo {volume}
  {A663}},\ \bibinfo {pages} {401} (\bibinfo {year} {2000}{\natexlab{b}})},\
  \Eprint {http://arxiv.org/abs/nucl-th/9908045} {arXiv:nucl-th/9908045
  [nucl-th]} \BibitemShut {NoStop}%
\bibitem [{\citenamefont {Bhagwat}\ and\ \citenamefont
  {Maris}(2008)}]{Bhagwat:2006pu}%
  \BibitemOpen
  \bibfield  {author} {\bibinfo {author} {\bibfnamefont {M.~S.}\ \bibnamefont
  {Bhagwat}}\ and\ \bibinfo {author} {\bibfnamefont {P.}~\bibnamefont
  {Maris}},\ }\href {\doibase 10.1103/PhysRevC.77.025203} {\bibfield  {journal}
  {\bibinfo  {journal} {Phys. Rev.}\ }\textbf {\bibinfo {volume} {C77}},\
  \bibinfo {pages} {025203} (\bibinfo {year} {2008})},\ \Eprint
  {http://arxiv.org/abs/nucl-th/0612069} {arXiv:nucl-th/0612069 [nucl-th]}
  \BibitemShut {NoStop}%
\bibitem [{\citenamefont {Goecke}\ \emph {et~al.}(2011)\citenamefont {Goecke},
  \citenamefont {Fischer},\ and\ \citenamefont {Williams}}]{Goecke:2010if}%
  \BibitemOpen
  \bibfield  {author} {\bibinfo {author} {\bibfnamefont {T.}~\bibnamefont
  {Goecke}}, \bibinfo {author} {\bibfnamefont {C.~S.}\ \bibnamefont {Fischer}},
  \ and\ \bibinfo {author} {\bibfnamefont {R.}~\bibnamefont {Williams}},\
  }\href {\doibase 10.1103/PhysRevD.83.094006, 10.1103/PhysRevD.86.099901}
  {\bibfield  {journal} {\bibinfo  {journal} {Phys. Rev.}\ }\textbf {\bibinfo
  {volume} {D83}},\ \bibinfo {pages} {094006} (\bibinfo {year} {2011})},\
  \bibinfo {note} {[Erratum: Phys. Rev.D86,099901(2012)]},\ \Eprint
  {http://arxiv.org/abs/1012.3886} {arXiv:1012.3886 [hep-ph]} \BibitemShut
  {NoStop}%
\bibitem [{\citenamefont {Ball}\ and\ \citenamefont
  {Chiu}(1980{\natexlab{a}})}]{Ball:1980ay}%
  \BibitemOpen
  \bibfield  {author} {\bibinfo {author} {\bibfnamefont {J.~S.}\ \bibnamefont
  {Ball}}\ and\ \bibinfo {author} {\bibfnamefont {T.-W.}\ \bibnamefont
  {Chiu}},\ }\href {\doibase 10.1103/PhysRevD.22.2542} {\bibfield  {journal}
  {\bibinfo  {journal} {Phys. Rev.}\ }\textbf {\bibinfo {volume} {D22}},\
  \bibinfo {pages} {2542} (\bibinfo {year} {1980}{\natexlab{a}})}\BibitemShut
  {NoStop}%
\bibitem [{\citenamefont {Ball}\ and\ \citenamefont
  {Chiu}(1980{\natexlab{b}})}]{Ball:1980ax}%
  \BibitemOpen
  \bibfield  {author} {\bibinfo {author} {\bibfnamefont {J.~S.}\ \bibnamefont
  {Ball}}\ and\ \bibinfo {author} {\bibfnamefont {T.-W.}\ \bibnamefont
  {Chiu}},\ }\href {\doibase 10.1103/PhysRevD.22.2550,
  10.1103/PhysRevD.23.3085} {\bibfield  {journal} {\bibinfo  {journal} {Phys.
  Rev.}\ }\textbf {\bibinfo {volume} {D22}},\ \bibinfo {pages} {2550} (\bibinfo
  {year} {1980}{\natexlab{b}})},\ \bibinfo {note} {[Erratum: Phys.
  Rev.D23,3085(1981)]}\BibitemShut {NoStop}%
\bibitem [{\citenamefont {Weil}\ \emph {et~al.}(2017)\citenamefont {Weil},
  \citenamefont {Eichmann}, \citenamefont {Fischer},\ and\ \citenamefont
  {Williams}}]{Weil:2017knt}%
  \BibitemOpen
  \bibfield  {author} {\bibinfo {author} {\bibfnamefont {E.}~\bibnamefont
  {Weil}}, \bibinfo {author} {\bibfnamefont {G.}~\bibnamefont {Eichmann}},
  \bibinfo {author} {\bibfnamefont {C.~S.}\ \bibnamefont {Fischer}}, \ and\
  \bibinfo {author} {\bibfnamefont {R.}~\bibnamefont {Williams}},\ }\href
  {\doibase 10.1103/PhysRevD.96.014021} {\bibfield  {journal} {\bibinfo
  {journal} {Phys. Rev.}\ }\textbf {\bibinfo {volume} {D96}},\ \bibinfo {pages}
  {014021} (\bibinfo {year} {2017})},\ \Eprint
  {http://arxiv.org/abs/1704.06046} {arXiv:1704.06046 [hep-ph]} \BibitemShut
  {NoStop}%
\end{thebibliography}%

\end{document}